\documentclass[a4paper,12pt,onecolumn]{article}
\usepackage{mathrsfs}
\usepackage{epsf}
\usepackage{graphicx}
\usepackage{latexsym}
\usepackage{CJK}
\usepackage{indentfirst}
\usepackage[colorlinks]{hyperref}
\usepackage {amsmath}
\usepackage {amsxtra}
\usepackage {amsbsy}
\usepackage {amscd}

\usepackage {graphicx}
\usepackage {graphics}
\usepackage {amssymb}
\usepackage {amsfonts}
\usepackage {float}
\usepackage {indentfirst}
\usepackage {a4wide}

\usepackage{subfigure}
\usepackage{color}

\allowdisplaybreaks

\newcommand{\be}{\begin{equation}}
\newcommand{\ee}{\end{equation}}
\newcommand{\ba}{\begin{array}{c}}
\newcommand{\ea}{\end{array}}
\newcommand{\bqa}{\begin{eqnarray}}
\newcommand{\eqa}{\end{eqnarray}}
\newcommand{\bqaa}{\begin{eqnarray*}}
\newcommand{\eqaa}{\end{eqnarray*}}

\begin{document}

\newcommand{\supercite}[1]{\textsuperscript{\cite{#1}}}
\frenchspacing \linespread{1.2}

\title{Pole Analysis of Unitarized  One Loop
$\chi$PT Amplitudes -- A Triple Channel Study }

\author{L.Y.~Dai$^{1}$ \footnote{Email:~\href{daily03@pku.edu.cn}{daily03@pku.edu.cn}},
        X.G.~Wang $^{2}$ \footnote{Email:~\href{wangxg84@ihep.ac.cn}{wangxg84@ihep.ac.cn}},
        H.Q.~Zheng$^{1}$ \footnote{Email:~\href{zhenghq@pku.edu.cn}{zhenghq@pku.edu.cn}}\\
{\small{$^1$ \it Department of Physics and State Key Laboratory of Nuclear Physics and}}\\
 {\small{\it Technology, Peking University, Beijing 100871, China}}\\
{\small{$^2$ \it Institute of High Energy Physics, Chinese Academy of Science, Beijing 100049, P.~R.~China}}
}

\date{September, 2010}

\maketitle

\begin{abstract}
In a previous paper~\cite{Dai11}, we proposed a method to
distinguish poles of different dynamical origin, in a unitarized
amplitude of $\pi\pi\ ,K\bar K$ system. That is based on the
observation that `A Breit-Wigner resonance should exhibit two poles
on different Riemann sheets which meet each other on the real axis
when $N_c=\infty$'. In this paper, we extend our previous
work~\cite{Dai11} to the $\pi\pi$-$K\bar K$-$\eta\eta$ three
channel  system. We reconfirm most of the predictions of
Ref.~\cite{Dai11}. Especially the $f_0(980)$ is of  $K\bar K$
molecule nature. Other poles, including the $\sigma$, are of
Breit--Wigner type.

\end{abstract}

\vspace{1cm}
\section{Introduction}
Hadron physics in the intermediate energy region (i.e., $\sim
0.5-2$GeV) is well known to be a difficult field of study, since
neither pQCD nor chiral perturbation theory (ChPT) is applicable
there. Especially the study on the property of light scalar mesons
has a long history of controversy.  Dispersion technique is found to
be powerful, in firmly establishing the existence of light scalar
mesons such as $\sigma$, $\kappa$ and in determining rather
precisely their pole locations~\cite{Zheng01}-\cite{KPY}. However,
the property of the light scalar mesons
($f_0(980)$,$a_0(980)$,$f_0(600)$, $K_0^*(700)$) remains   somewhat
mysterious.~\footnote{This problem is discussed, for example, in
Ref.~\cite{zhengcd09}.}

One commonly used method in the literature to attack this problem is
the so called ``Inverse Amplitude Method`` or ``chiral unitarization
approach``. The $\pi\pi-K\bar{K}$ and $\pi\eta-K\bar{K}$ couple
channel system is extensively studied~\cite{OOP98}-\cite{Guo1112},
giving the sheet II trajectory of light
mesons~\cite{Pelaez02}-\cite{Uehara04} below 1GeV. In
Ref.~\cite{Dai11} it is pointed out that a couple channel
Breit--Wigner resonance should exhibit two poles on different
Riemann sheets and reach the same position on the real axis when
$N_c=\infty$.  In this point of view `searching for accompanying
shadow pole` is stressed which is overlooked by most previous
studies. An advantage of this method is that the criteria is
unaltered when changing parameters and hence avoids parameter
dependence. The worry of being model (Pad\'e approximation)
dependent is also, at least partly, avoided, since one can examine
the pole structure of well known particles (such as $\rho$ and $K^*$
resonances) to justify the validity of using Pad\'e approximation in
studying analyticity property of the pole.  In this paper we follow
the idea of Ref.~\cite{Dai11} and extend it to the situation with
$\pi\pi\ ,K\bar K\ ,\eta\eta$ three channel system (together with
$\pi\eta\ ,K\bar K$ system when studying $a_0(980)$). We find that,
with $\eta\eta$ channel included,  qualitative results obtained in
Ref.~\cite{Dai11} remain valid. Especially, it is found that the
$f_0(980)$ pole remains to be a single trajectory. Nevertheless, we
 point out that a careless mistake occurs in Ref.~\cite{Dai11}: the
$\sigma$, $\kappa$ were mistakenly claimed to have only single
trajectory in two channel case. According to a careful re-analysis,
they both    contain a shadow pole on sheet III -- hence the
$\sigma$, $\kappa$ mesons are also of couple channel Breit -- Wigner
resonance origin.

This paper is organized as follows. In section 2 we give a brief
introduction to partial wave projection,  analytical continuation,
and Pad\'e unitarization.  In section 3 we fit the experimental
phase shift and inelasticity  to fix the low energy constants (LECs)
and extract pole positions.  In section 4 we analyze the trajectory
of the poles.  The last section is for conclusion.

\section{Unitarization}
\subsection{Partial wave expansion}
The ChPT scattering amplitudes have been given
in~\cite{Pelaez02,Meissner91} and  the iso-spin decomposed formulas
are given in~\cite{Dai11}: The projection in definite angular
momentum is given by
\begin{equation}
T^{(I,J)}=\frac{1}{32N\pi}\int_{-1}^{1}d(cos\theta)T^I(s,\cos\theta)P_J(cos\theta)\,
\end{equation}
where I is isospin, J denotes the total angular momentum.
 Here $N=2$ for $\pi\pi, \eta\eta\to \pi\pi, \eta\eta$ amplitudes,
$N=\sqrt{2}$ for $\pi\pi, \eta\eta\to K\bar K$ amplitudes, and N=1
for other cases. From now on we omit the I,J index for simplicity.

\subsection{Pad$\acute{e}$ approximation}
Matrix Pad\'e approximation was derived to obtain unitarized
amplitudes from the $\mathcal{O}(p^2)$ and $\mathcal{O}(p^4)$  ChPT
scattering amplitudes, which is equivalent to:
\begin{equation}
T=T^{(2)}\cdot[T^{(2)}-T^{(4)}]^{-1}\cdot T^{(2)}\ .
\end{equation}
For a three channel case, the unitarized amplitudes satisfy such unitarity conditions,
\begin{eqnarray}\label{unitarity}
\mathrm{Im}_R T_{11}&=&T_{11}\rho_1 T_{11}^{*}\theta(s-4m_{\pi}^2)+T_{12}\rho_2 T_{12}^{*}\theta(s-4m_{K}^2)+T_{13}\rho_3 T_{13}^{*}\theta(s-4m_{\eta}^2)\ ,\nonumber\\
\mathrm{Im}_R T_{12}&=&T_{11}\rho_1 T_{12}^{*}\theta(s-4m_{\pi}^2)+T_{12}\rho_2 T_{22}^{*}\theta(s-4m_{K}^2)+T_{13}\rho_3 T_{23}^{*}\theta(s-4m_{\eta}^2)\ ,\nonumber\\
\mathrm{Im}_R T_{13}&=&T_{11}\rho_1 T_{13}^{*}\theta(s-4m_{\pi}^2)+T_{12}\rho_2 T_{23}^{*}\theta(s-4m_{K}^2)+T_{13}\rho_3 T_{33}^{*}\theta(s-4m_{\eta}^2)\ ,\nonumber\\
\mathrm{Im}_R T_{22}&=&T_{12}\rho_1 T_{12}^{*}\theta(s-4m_{\pi}^2)+T_{22}\rho_2 T_{22}^{*}\theta(s-4m_{K}^2)+T_{23}\rho_3 T_{23}^{*}\theta(s-4m_{\eta}^2)\ ,\nonumber\\
\mathrm{Im}_R T_{23}&=&T_{12}\rho_1 T_{13}^{*}\theta(s-4m_{\pi}^2)+T_{22}\rho_2 T_{23}^{*}\theta(s-4m_{K}^2)+T_{23}\rho_3 T_{33}^{*}\theta(s-4m_{\eta}^2)\ ,\nonumber\\
\mathrm{Im}_R T_{33}&=&T_{13}\rho_1 T_{13}^{*}\theta(s-4m_{\pi}^2)+T_{23}\rho_2 T_{23}^{*}\theta(s-4m_{K}^2)+T_{33}\rho_3 T_{33}^{*}\theta(s-4m_{\eta}^2)\ .
\end{eqnarray}
Here the subscript 1 represents  $\pi\pi$, 2 for $K\bar K$, and 3
represents $\eta\eta$.  The fourth equation only holds true above
$4m_K^2-4m_{\pi}^2$~\cite{Kennedy62} and the sixth equation only
holds true above $4m_{\eta}^2-4m_{\pi}^2$. The same as in couple
channel, it violates unitarity as the left hand cut $(-\infty,
4m_{\eta}^2-4m_{\pi}^2]$ appears not only in $T_{33}$, but also in
the other five T matrix elements.

 Partial wave $S$ matrix elements are given by
\begin{eqnarray}
S_{11}&=&1+2i\rho_1(s)T_{11}(s)\ ,\nonumber\\
S_{12}&=&2i\sqrt{\rho_{1}(s)\rho_2(s)}T_{12}(s)\,\nonumber\\
S_{13}&=&2i\sqrt{\rho_{1}(s)\rho_3(s)}T_{13}(s)\,\nonumber\\
S_{22}&=&1+2i\rho_2(s)T_{22}(s)\ ,\nonumber\\
S_{23}&=&2i\sqrt{\rho_{2}(s)\rho_3(s)}T_{23}(s)\,\nonumber\\
S_{33}&=&1+2i\rho_3(s)T_{33}(s)\ .
\end{eqnarray}
The $3\times3$ S-matrix parametrization  is given in the
paper~\cite{Badalyan82,Guo1112}:
\begin{equation}
\Phi(x)=\left(
          \begin{array}{ccc}
   \eta_{11} \exp^{2i\delta_{11}}                     & \eta_{12} \exp^{i \delta_{12}}       & \eta_{13} \exp^{i\delta_{13}}  \\
   \eta_{12} \exp^{i\delta_{12}}                      & \eta_{22} \exp^{2i\delta_{22}}       & \eta_{23} \exp^{i\delta_{23}}   \\
   \eta_{13} \exp^{i\delta_{13}}                      & \eta_{23} \exp^{2i\delta_{23}}       & \eta_{33} \exp^{2i\delta_{33}} \\
          \end{array}
        \right)\ .
\end{equation}

\subsection{Analytical Continuation}
To extend the scattering amplitudes in the complex plane, we need
analytic continuation. The multi-channel continuation has been given
by the paper according to the cuts caused by phase space factor
$\rho$~\cite{Krupa95}. Here we use the following order of analytical
continuation for simplicity:
\begin{eqnarray}
\mathrm{The\,\,\, first\,\,\, continuation:}  &T^{I}(s-i\epsilon)=T^{II}(s+i\epsilon),  & \nonumber\\
\mathrm{The\,\,\, second\,\,\, continuation:} &T^{I}(s-i\epsilon)=T^{III}(s+i\epsilon), & T^{II}(s-i\epsilon)=T^{V}(s+i\epsilon),\nonumber\\
\mathrm{The\,\,\, third\,\,\, continuation:}   &T^{I}(s-i\epsilon)=T^{IV}(s+i\epsilon),  & T^{II}(s-i\epsilon)=T^{VIII}(s+i\epsilon),\nonumber\\
                         &T^{III}(s-i\epsilon)=T^{VI}(s+i\epsilon),& T^{V}(s-i\epsilon)=T^{VII}(s+i\epsilon)\ .\nonumber
\end{eqnarray}
They generate the Riemann sheets:
\begin{table}[h]
\begin{center}
\begin{tabular}{c c c c c c c c c}
\hline
               & I    & II  & III  & IV  & V  & VI  & VII  & VIII   \\
\hline \hline
$\rho_1$       & +    & -   &   -  &  -  & +  &  +  &   -  & +      \\
$\rho_2$       & +    & +   &   -  &  -  & -  &  +  &   +  & -      \\
$\rho_3$       & +    & +   &   +  &  -  & +  &  -  &   -  & -      \\
\hline
\end{tabular}
\caption{\label{continuation}The definition of Riemann sheets. }
\end{center}
\end{table}

\section{Fixing LECs and pole trajectory}
\subsection{Fit of all channels}
\begin{figure}[t]
\begin{minipage}[p]{0.31\textwidth}
\centering \subfigure[ ]{
\includegraphics[width=0.86\textwidth]{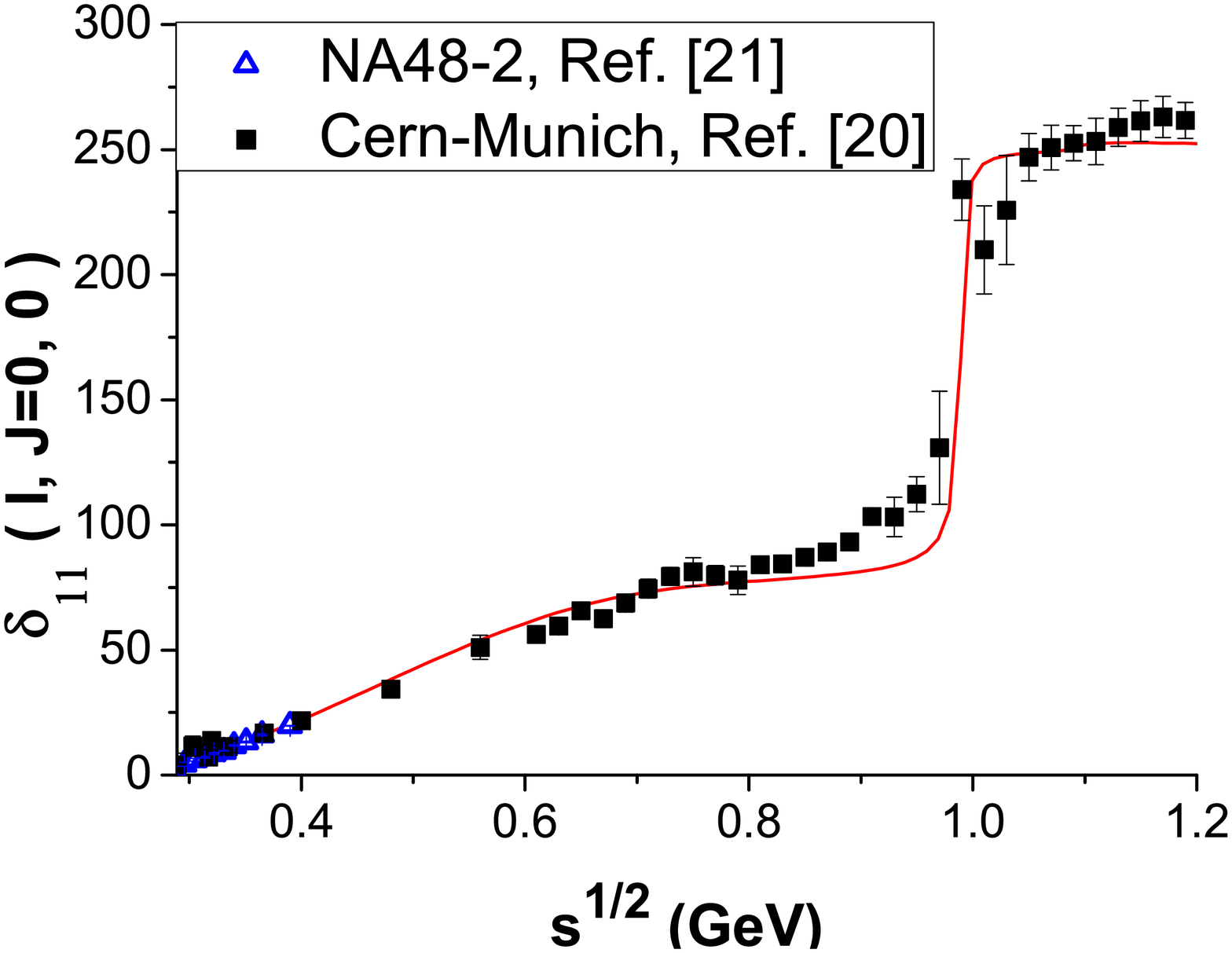}}
\end{minipage}
\begin{minipage}[p]{0.31\textwidth}
\centering \subfigure[ ]{
\includegraphics[width=0.86\textwidth]{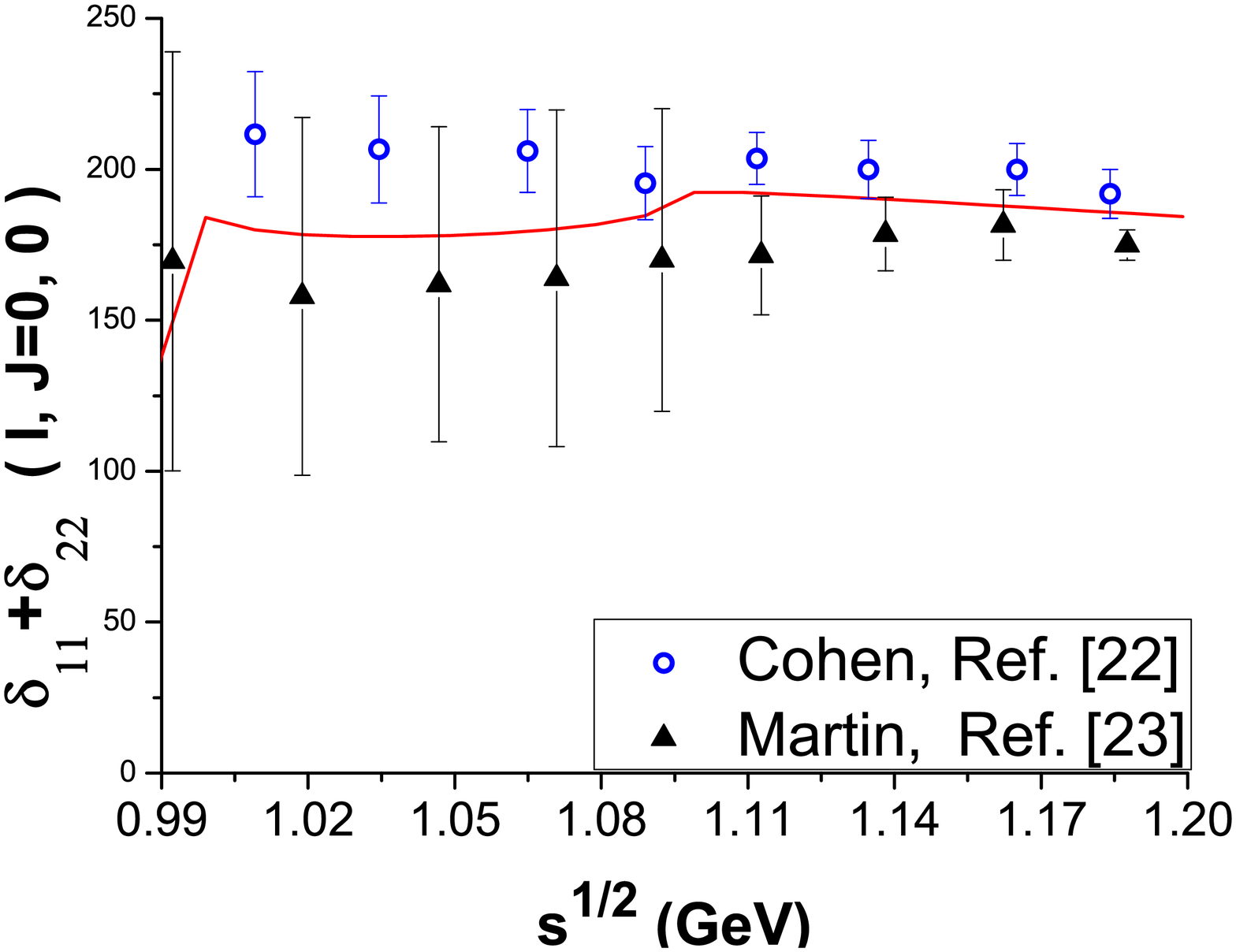}}
\end{minipage}
\begin{minipage}[p]{0.31\textwidth}
\centering \subfigure[ ]{
\includegraphics[width=0.86\textwidth]{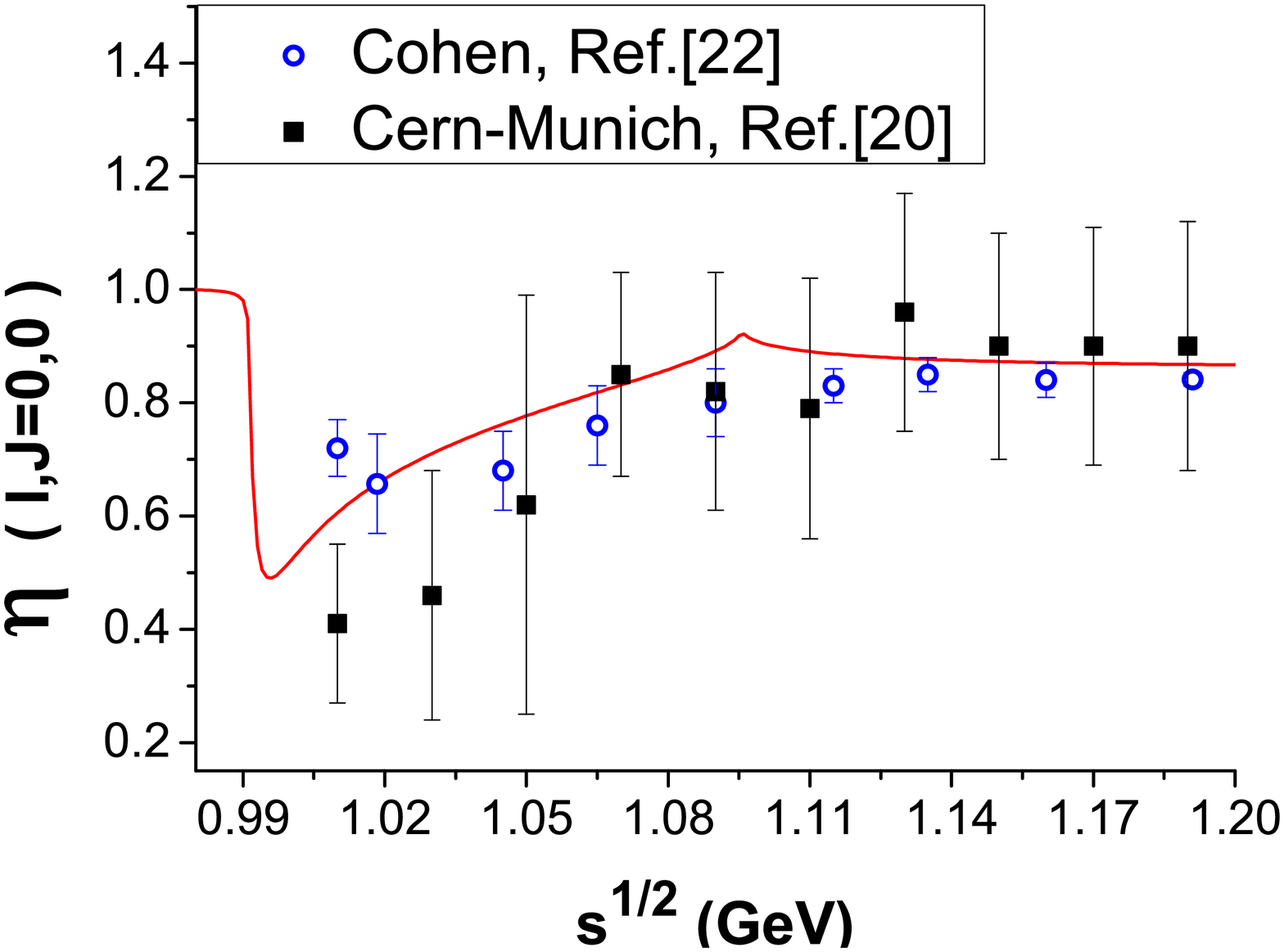}}
\end{minipage}
\begin{minipage}[p]{0.31\textwidth}
\centering \subfigure[ ]{
\includegraphics[width=0.86\textwidth]{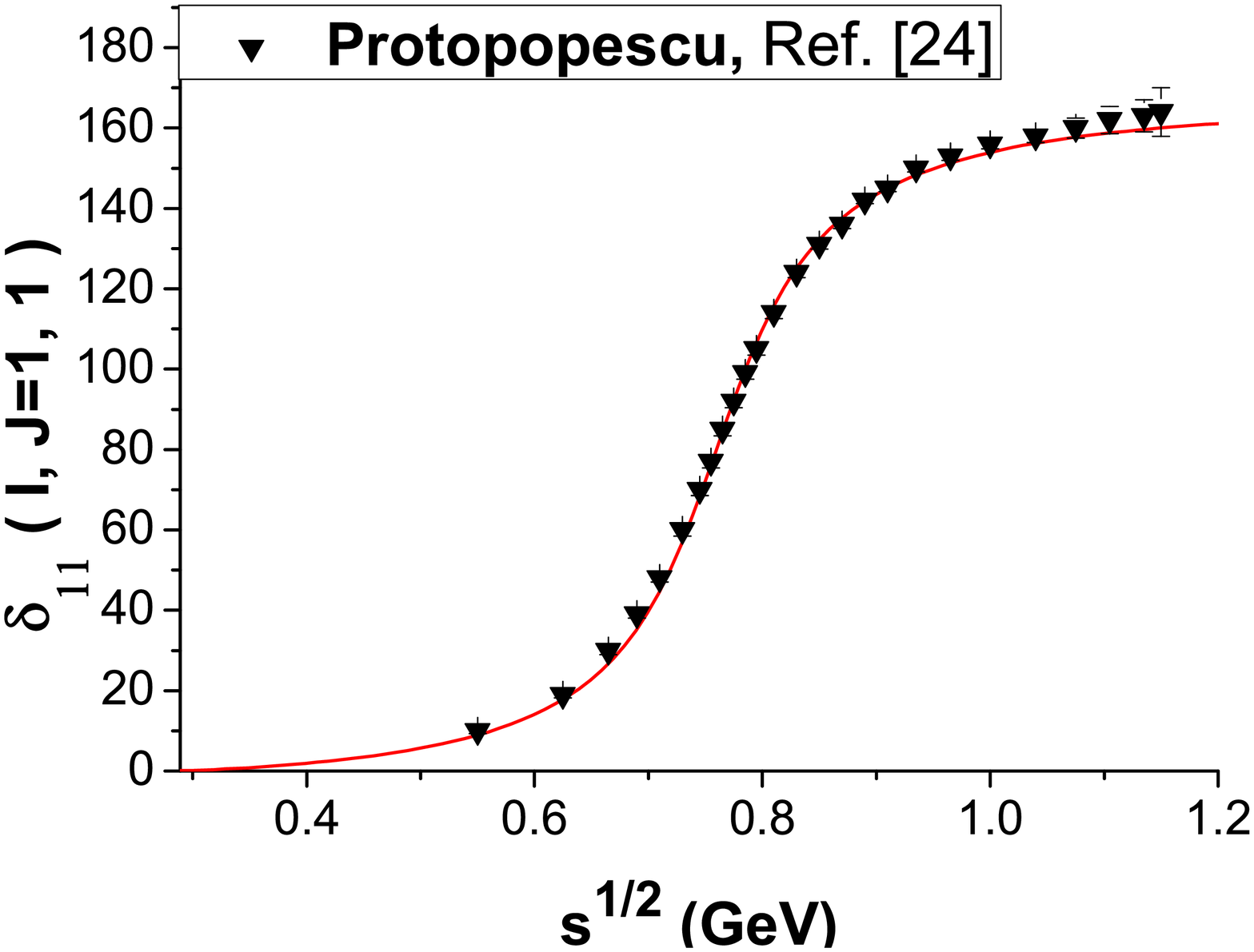}}
\end{minipage}
\begin{minipage}[p]{0.31\textwidth}
\centering \subfigure[ ]{
\includegraphics[width=0.86\textwidth]{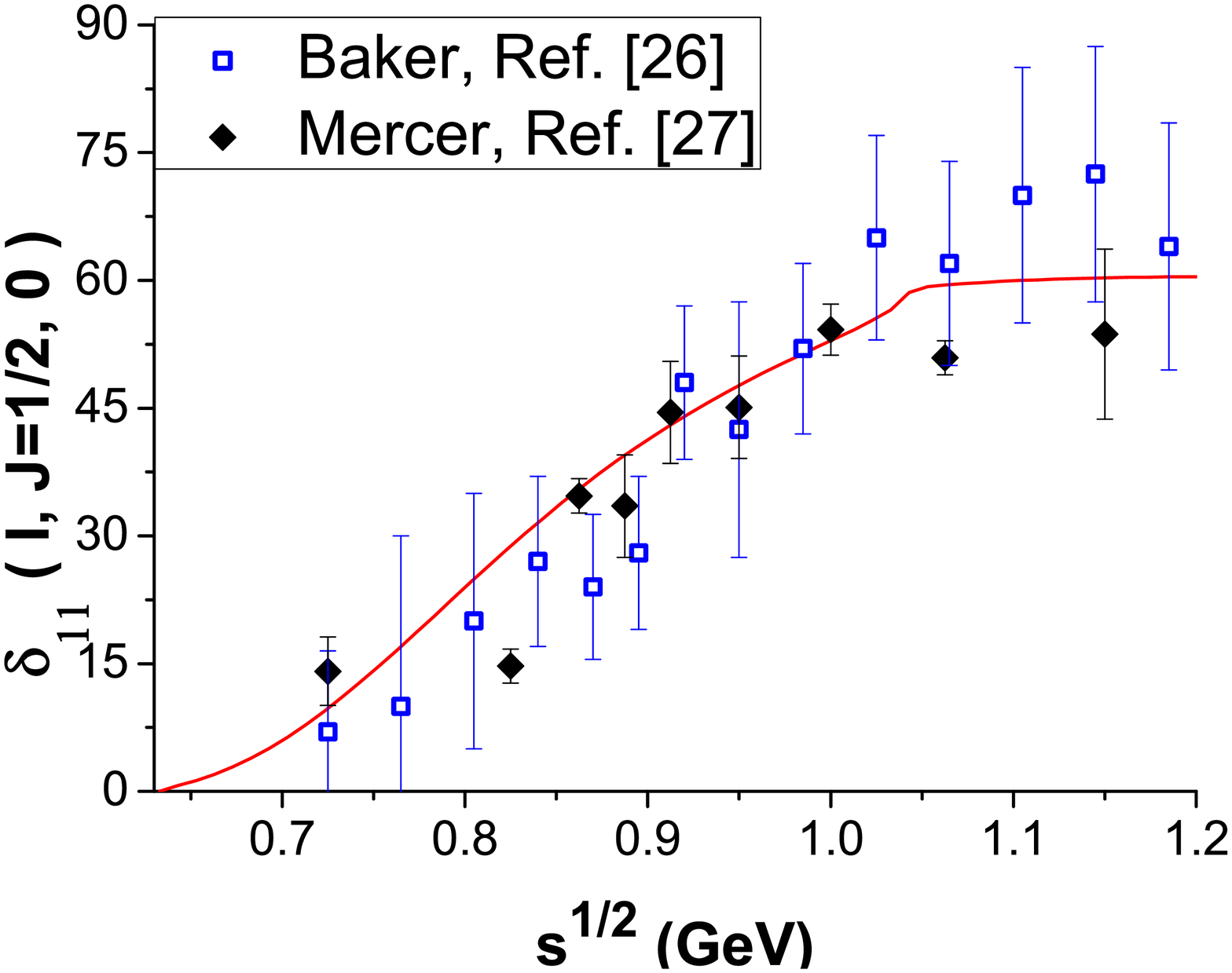}}
\end{minipage}
\begin{minipage}[p]{0.31\textwidth}
\centering \subfigure[ ]{
\includegraphics[width=0.86\textwidth]{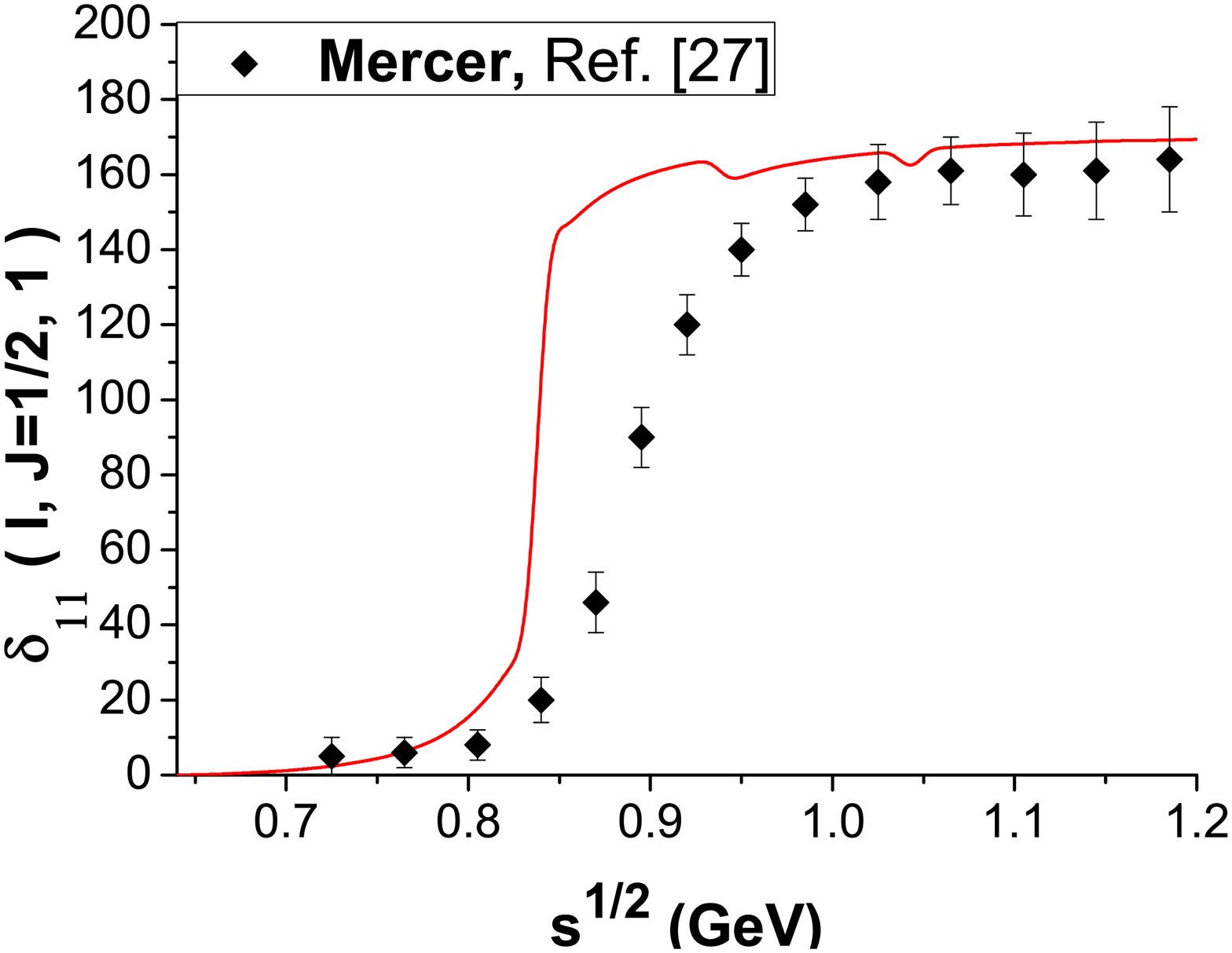}}
\end{minipage}
\begin{minipage}[p]{0.31\textwidth}
\centering \subfigure[ ]{
\includegraphics[width=0.86\textwidth]{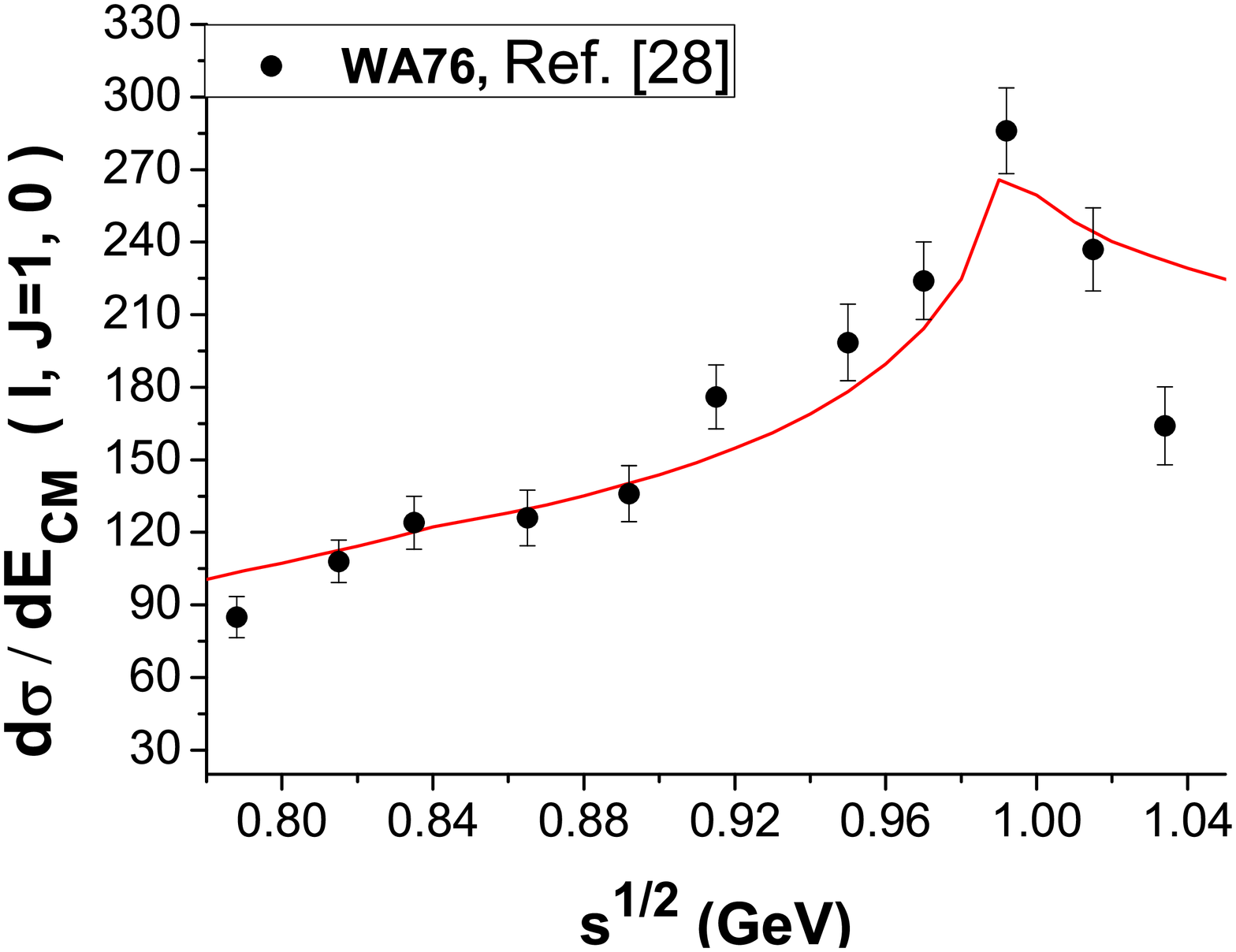}}
\end{minipage}
\begin{minipage}[p]{0.31\textwidth}
\centering \subfigure[ ]{
\includegraphics[width=0.86\textwidth]{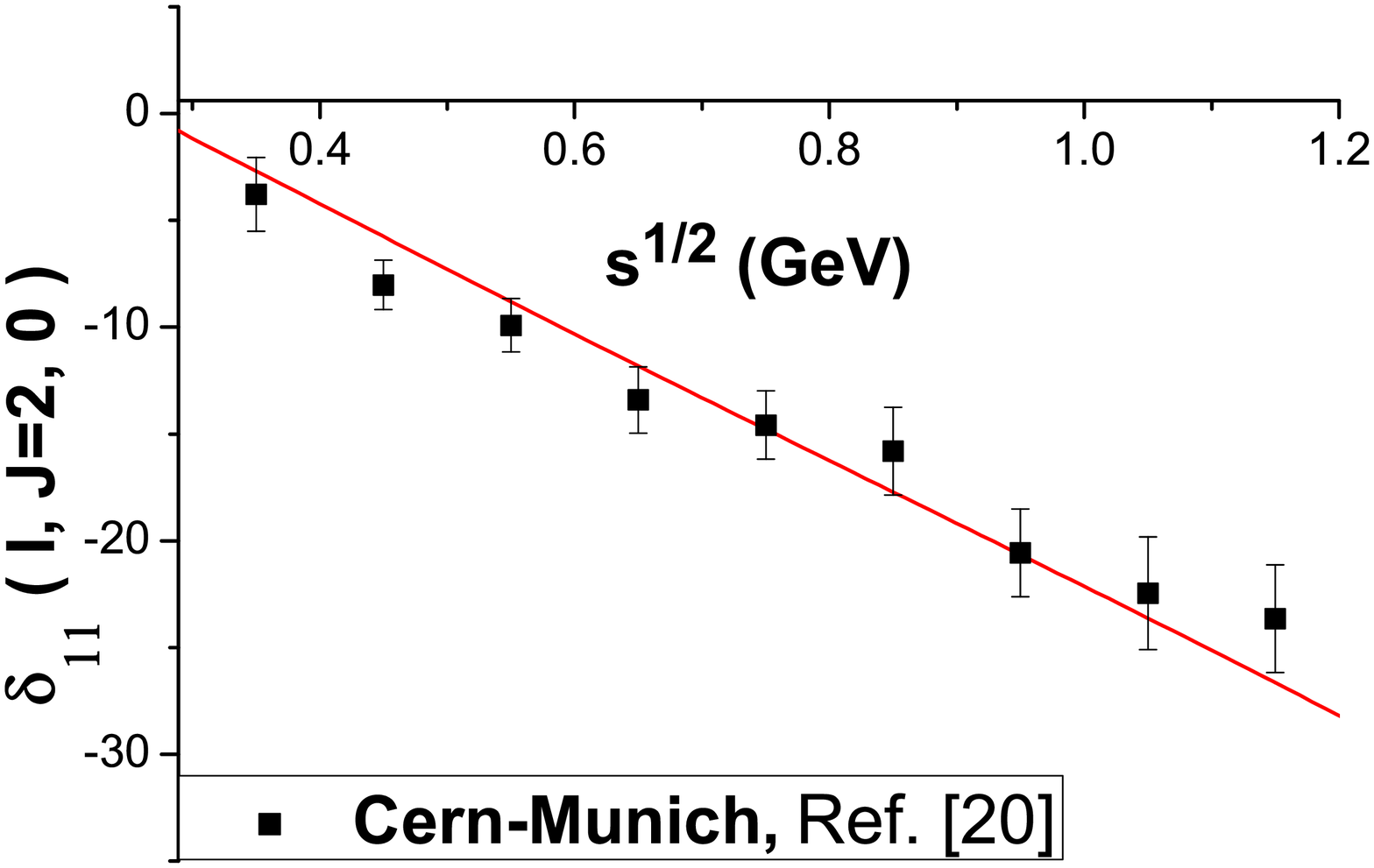}}
\end{minipage}
\begin{minipage}[p]{0.31\textwidth}
\centering \subfigure[ ]{
\includegraphics[width=0.86\textwidth]{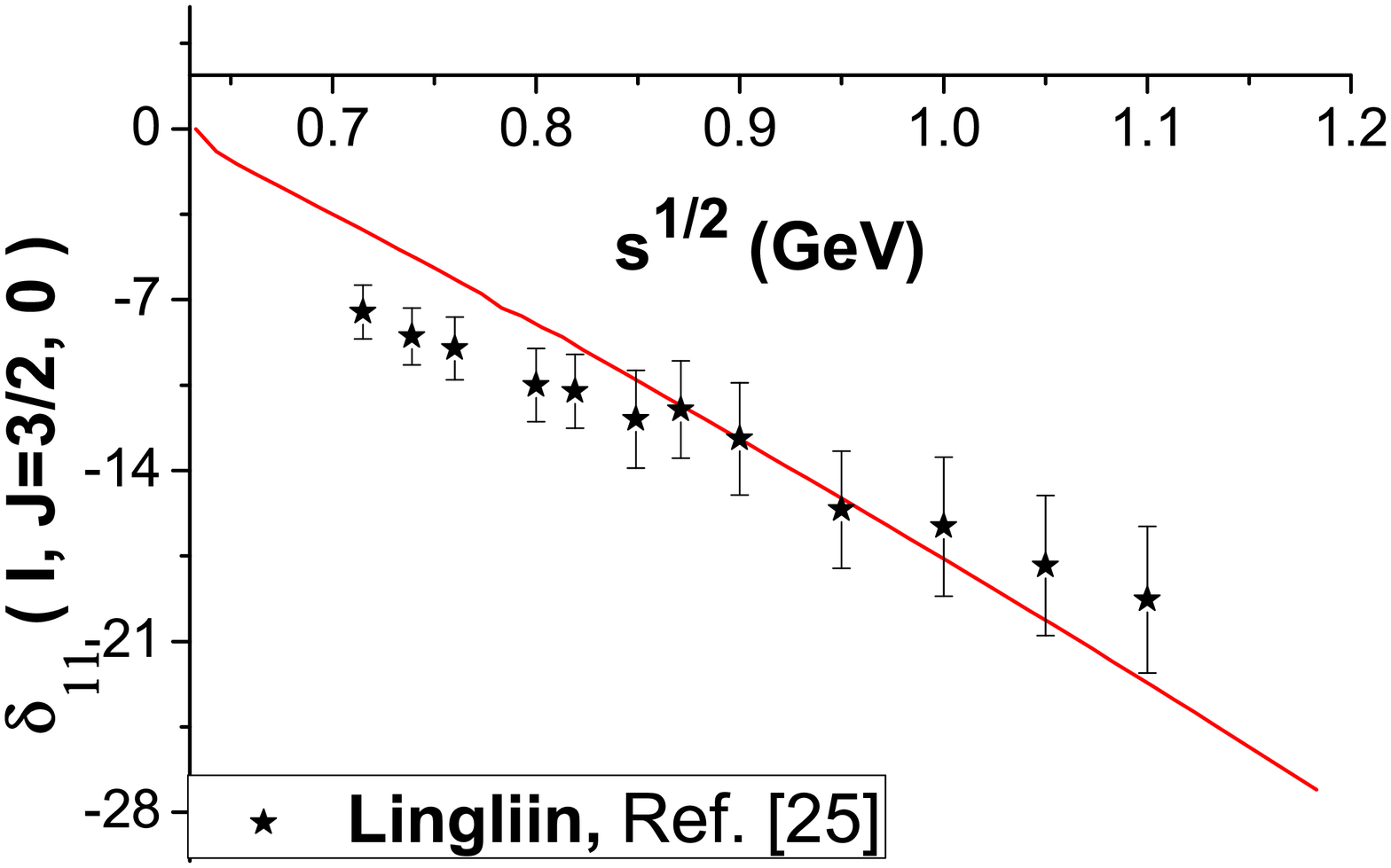}}
\end{minipage}
\caption{\label{phase}  Fit phase shift, inelasticity and cross
section.}
\end{figure}

We fit data in $IJ=00$ $\pi\pi$-$K\bar{K}$-$\eta\eta$ triple
channel, $IJ=20$ $\pi\pi$ single channel, $IJ=\frac{3}{2}\,0$ $\pi
K$ single channel,  $IJ=11$  $\pi\pi$-$K\bar{K}$ couple channel,
$IJ=10$ $\pi\eta$-$K\bar{K}$ couple channel, and
$IJ=\frac{1}{2}\,0,\frac{1}{2}\,1$  $\pi K$-$\eta K$ couple channel.
The data we used are as follows: for $IJ=00$, the
$\pi\pi\rightarrow\pi\pi$ phase  is from~\cite{CERN-Munich,NA48},
$\pi\pi\rightarrow K\bar{K}$ phase from~\cite{eta-Cohen,Martin};
$IJ=11$ $\pi\pi\rightarrow\pi\pi$ phase from~\cite{Protopopescu};
$IJ=20$ $\pi\pi$ phase from~\cite{CERN-Munich}; $IJ=\frac{3}{2}\,0$
$\pi K$ phase from~\cite{Linglin1973};  and $IJ=\frac{1}{2}\,0$,
$IJ=\frac{1}{2}\,1$ $\pi K\rightarrow\pi K$ phase
from~\cite{Baker1975,Mercer1971}. For $IJ=10$ , we use the $\pi\eta$
effective mass distribution data from the $pp\rightarrow
p(\eta\pi^+\pi^-)p$ reaction studied by WA76
Collaboration~\cite{WA76}, with the background given
by~\cite{Flatte76}. Here we use the same spectrum function as that
of~\cite{Pelaez02},
\begin{equation}
\frac{d\sigma}{dE_{cm}}=c p_{\pi\eta}|T_{\pi\eta\rightarrow K\bar{K}}^{IJ=10}|^2+b.g.\ ,
\end{equation}
where $c$ is a normalization factor. In additional, the data we fit
are among the energy region $4m_{\pi}^2$ -- $1.2GeV$, which is much
higher than the validity domain of ChPT. Parameters other than LECs
are given by: $f_{\pi}=0.0924GeV$, $m_{\pi}=0.1373GeV$,
$m_{K}=0.4957GeV$, $m_{\eta}=0.5475GeV$, and the renormalization
scale of ChPT amplitudes is taken to be $\mu=0.7755GeV$. The
numerical results are shown in Table~\ref{LECs} and
Fig.~\ref{phase}. We also list the value of ChPT and results
obtained in Ref.~\cite{Dai11} in Table~\ref{LECs} for comparison. We
find that the LECs obtained from $IJ=00$
$\pi\pi$-$K\bar{K}$-$\eta\eta$ triple channel fit can also reproduce
the $\pi\pi$-$K\bar{K}$ couple channel data very well.
\begin{table}[h]
\begin{center}
\begin{tabular}{|c||c|c|c|}
\hline
                 &  Fit                                &  Ref.~\cite{Dai11}            & ChPT ($10^{-3}$)       \\
\hline \hline
$L_{1r}$         &      $1.18\pm0.01 $                 &  $1.25\pm 0.02$                 &     $0.4\pm 0.3$       \\
$L_{2r}$         &      $1.67\pm0.01  $                &  $1.63\pm 0.03$                 &     $1.4\pm 0.3$       \\
$L_{3r}$         &      $-3.91\pm0.01 $                &  $-4.07\pm 0.02$                &     $-3.5\pm 1.1$      \\
$L_{4r}$         &      $-0.22\pm0.01 $                &  $0.30\pm 0.02$                 &     $-0.3\pm 0.5$      \\
$L_{5r}$         &      $-0.99\pm0.08 $                &  $-1.44\pm 0.42$                &     $1.4\pm 0.5$       \\
$2L_{6r}+L_{8r}$ &      $-0.50\pm0.03 $                &  $0.32\pm 0.13$                 &     $0.5\pm0.7$        \\
$2L_{7r}+L_{8r}$ &      $2.26\pm0.15 $                 &  $1.38\pm 0.38$                 &     $0.5\pm0.7$        \\
\hline
\end{tabular}
\caption{\label{LECs}LECs obtained from fit.  }
\end{center}
\end{table}

\begin{figure}[p]
\begin{minipage}[p]{0.5\textwidth}
\centering \subfigure[ ]{
\includegraphics[width=0.9\textwidth]{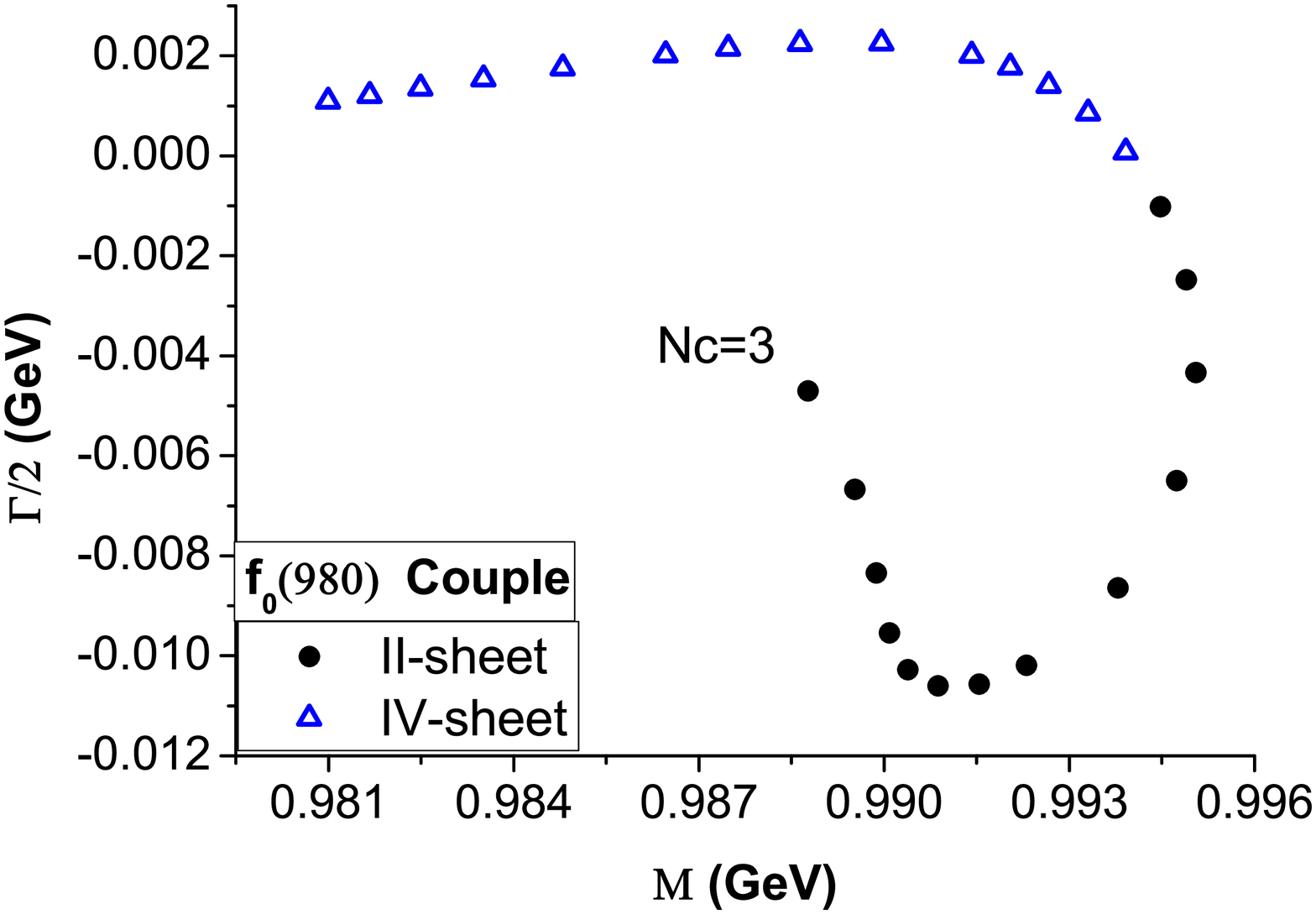}}
\end{minipage}
\begin{minipage}[p]{0.5\textwidth}
\centering \subfigure[ ]{
\includegraphics[width=0.9\textwidth]{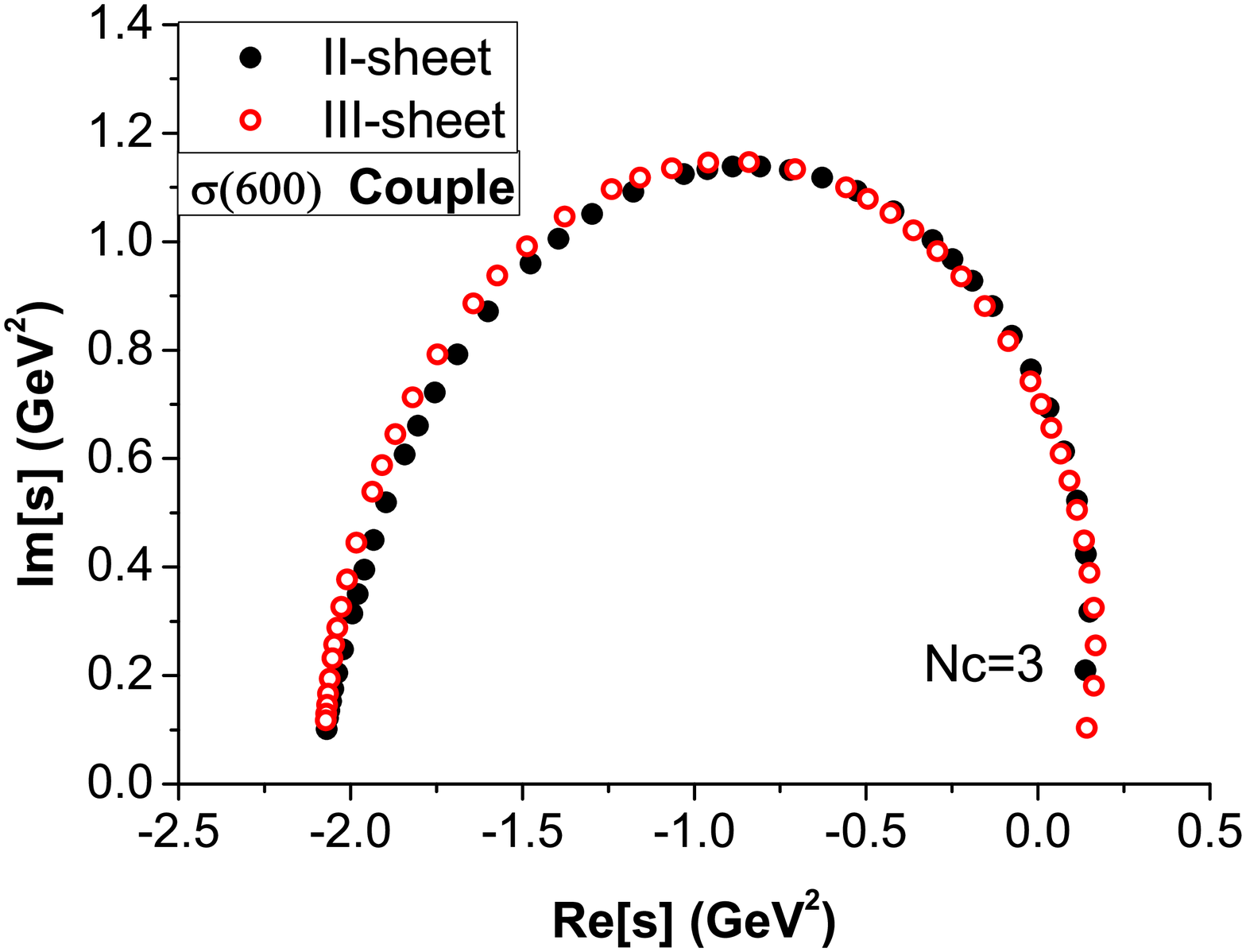}}
\end{minipage}
\begin{minipage}[p]{0.5\textwidth}
\centering \subfigure[ ]{
\includegraphics[width=0.9\textwidth]{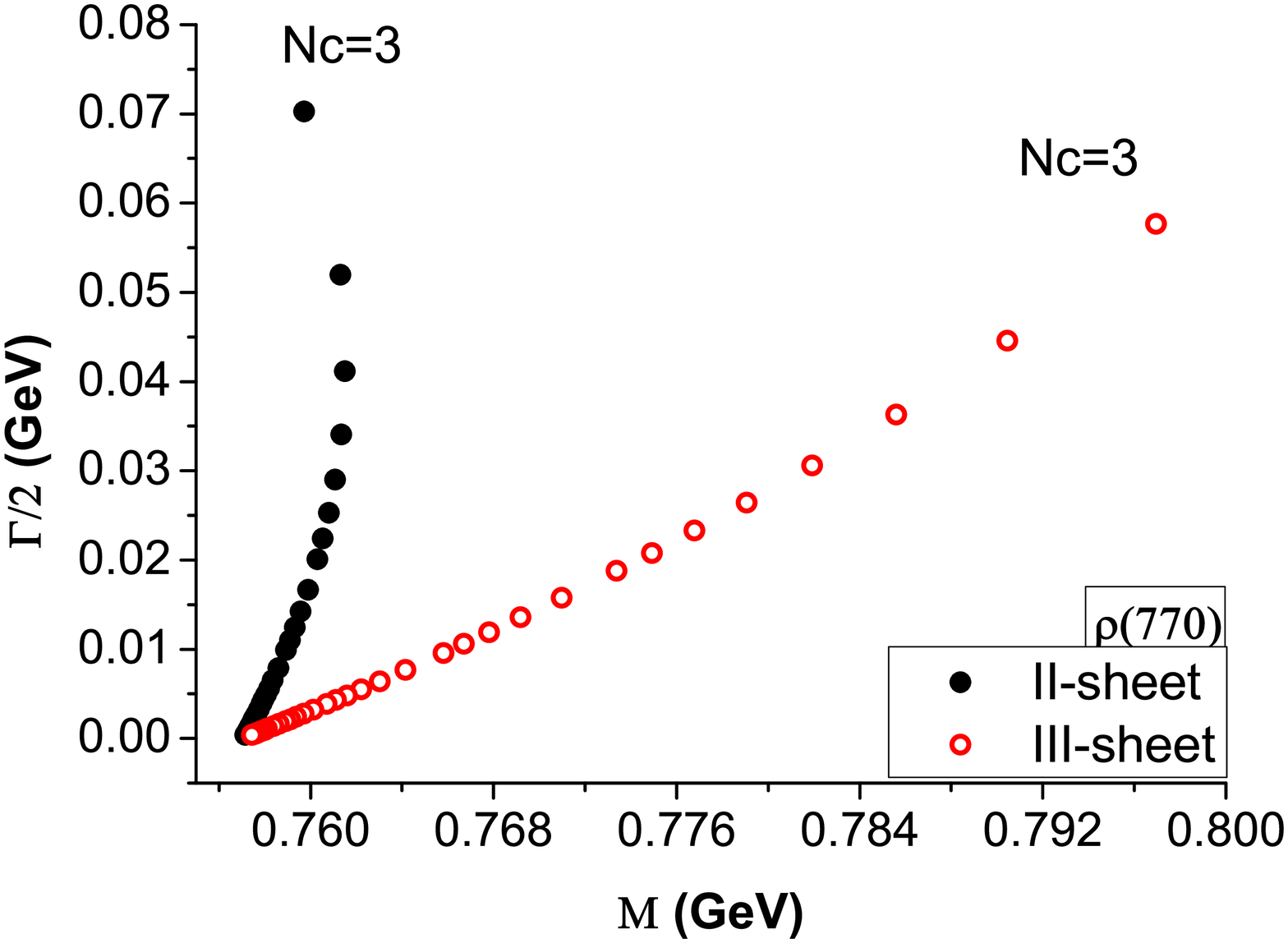}}
\end{minipage}
\begin{minipage}[p]{0.5\textwidth}
\centering \subfigure[ ]{
\includegraphics[width=0.9\textwidth]{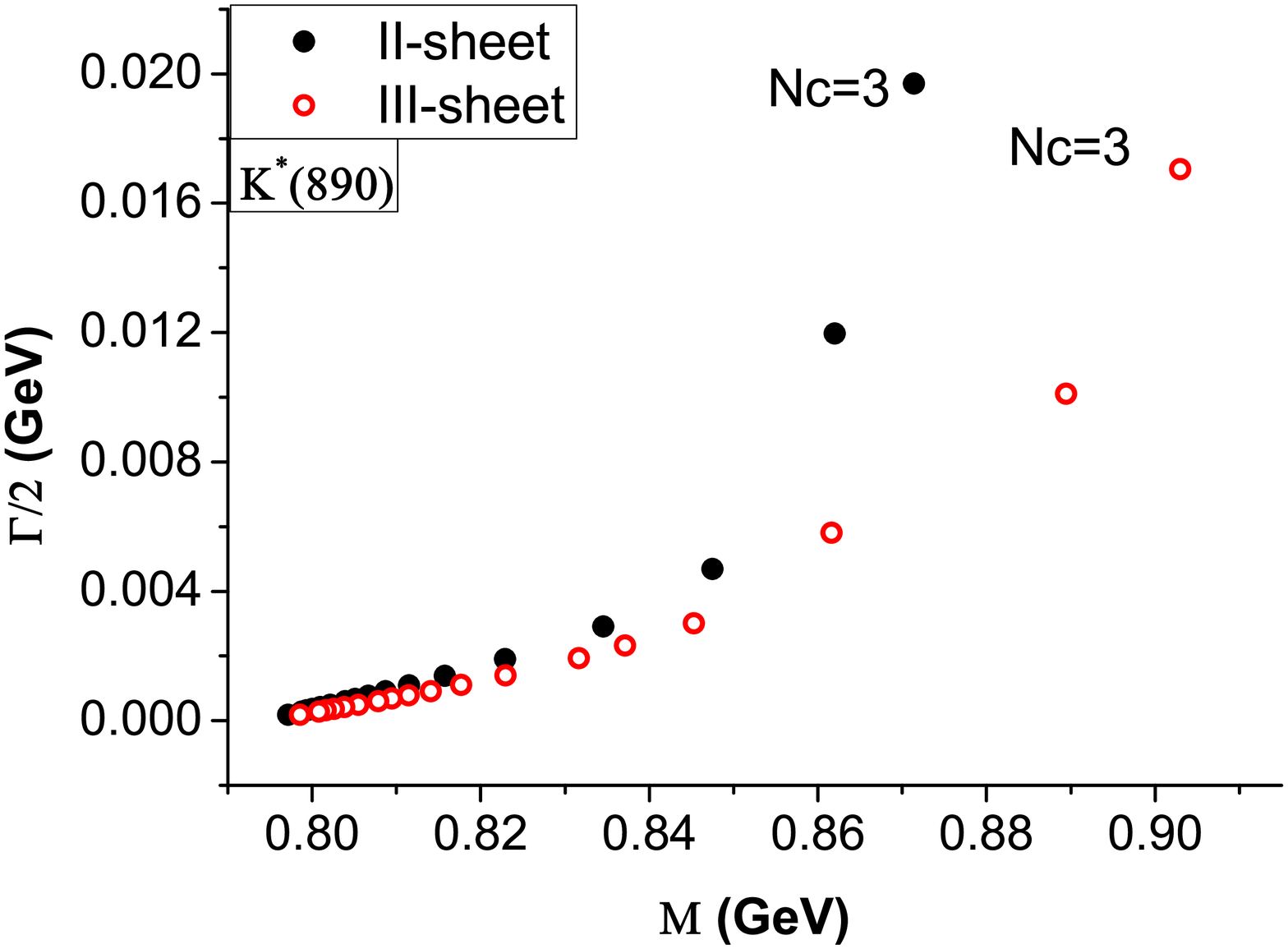}}
\end{minipage}
\begin{minipage}[p]{0.5\textwidth}
\centering \subfigure[ ]{
\includegraphics[width=0.9\textwidth]{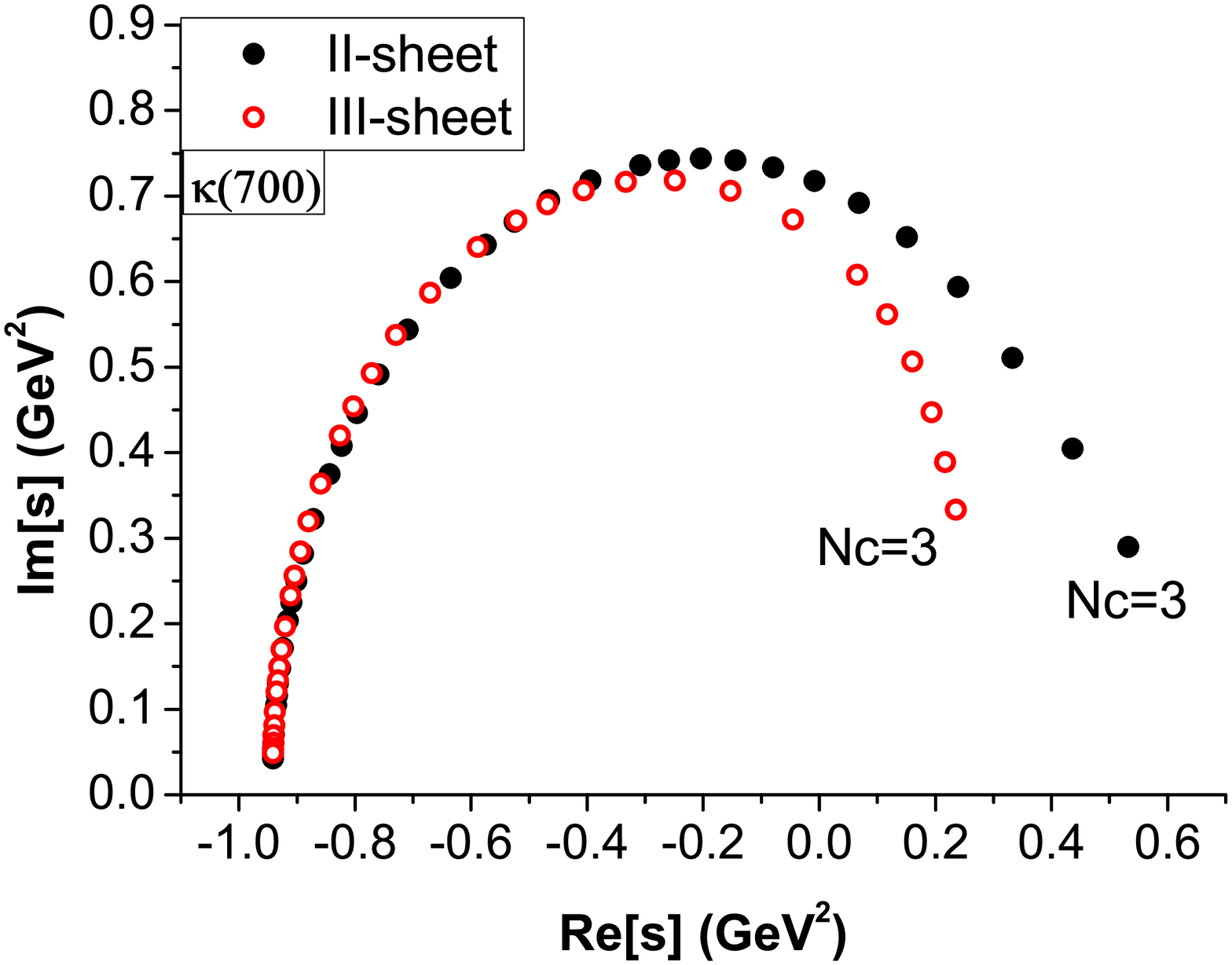}}
\end{minipage}
\begin{minipage}[p]{0.5\textwidth}
\centering \subfigure[ ]{
\includegraphics[width=0.9\textwidth]{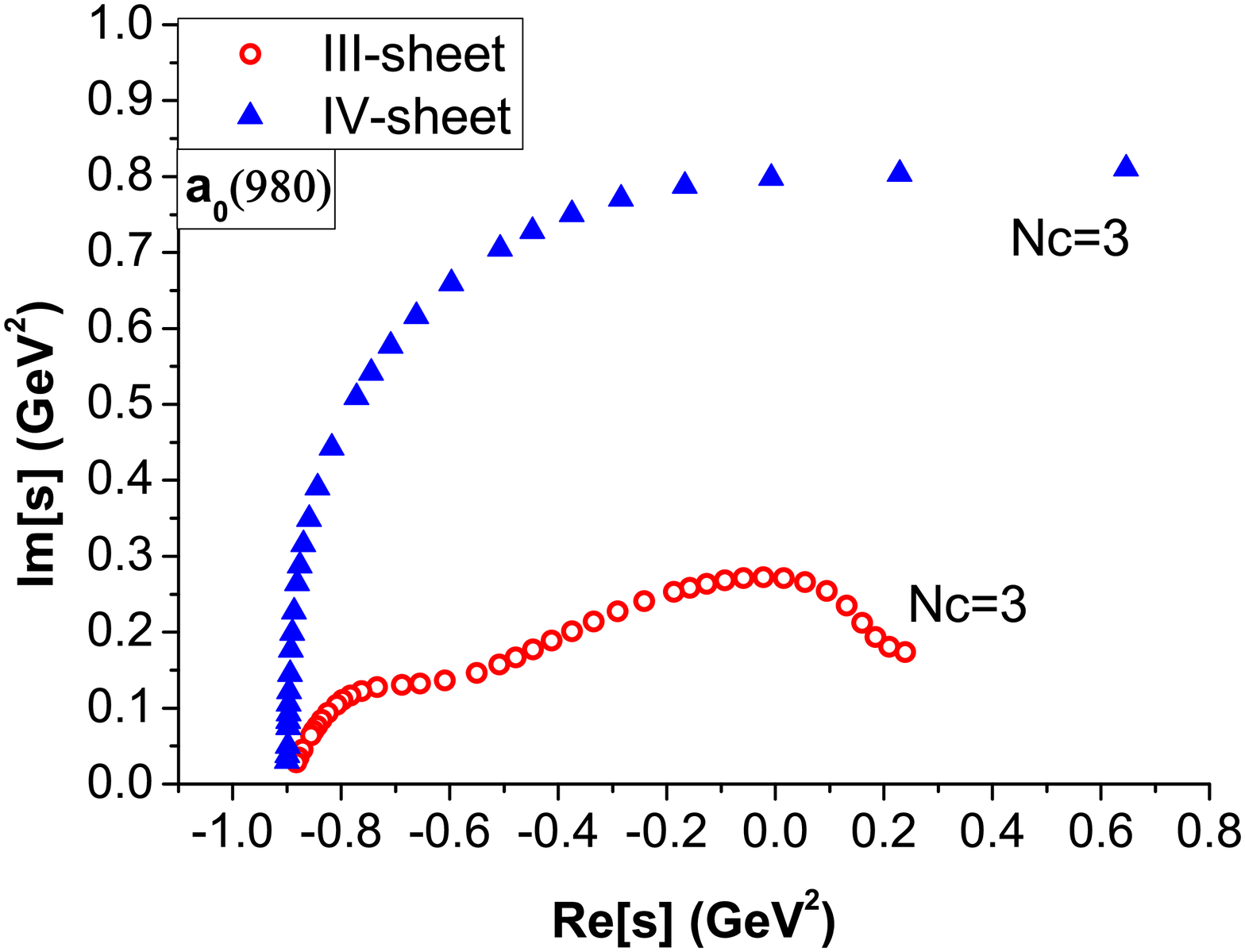}}
\end{minipage}
\begin{minipage}[p]{0.5\textwidth}
\centering \subfigure[ ]{
\includegraphics[width=0.9\textwidth]{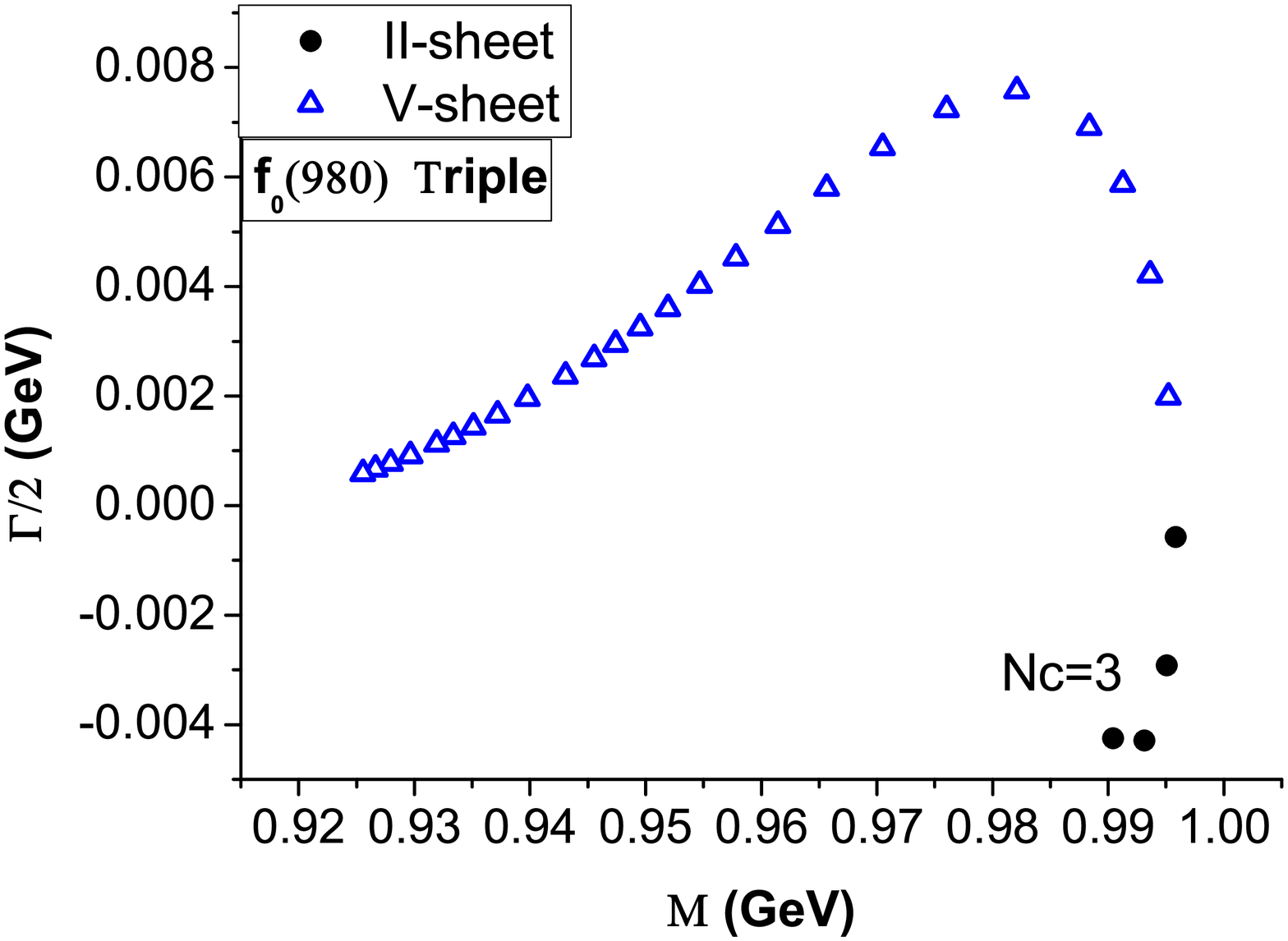}}
\end{minipage}
\begin{minipage}[p]{0.5\textwidth}
\centering \subfigure[ ]{
\includegraphics[width=0.9\textwidth]{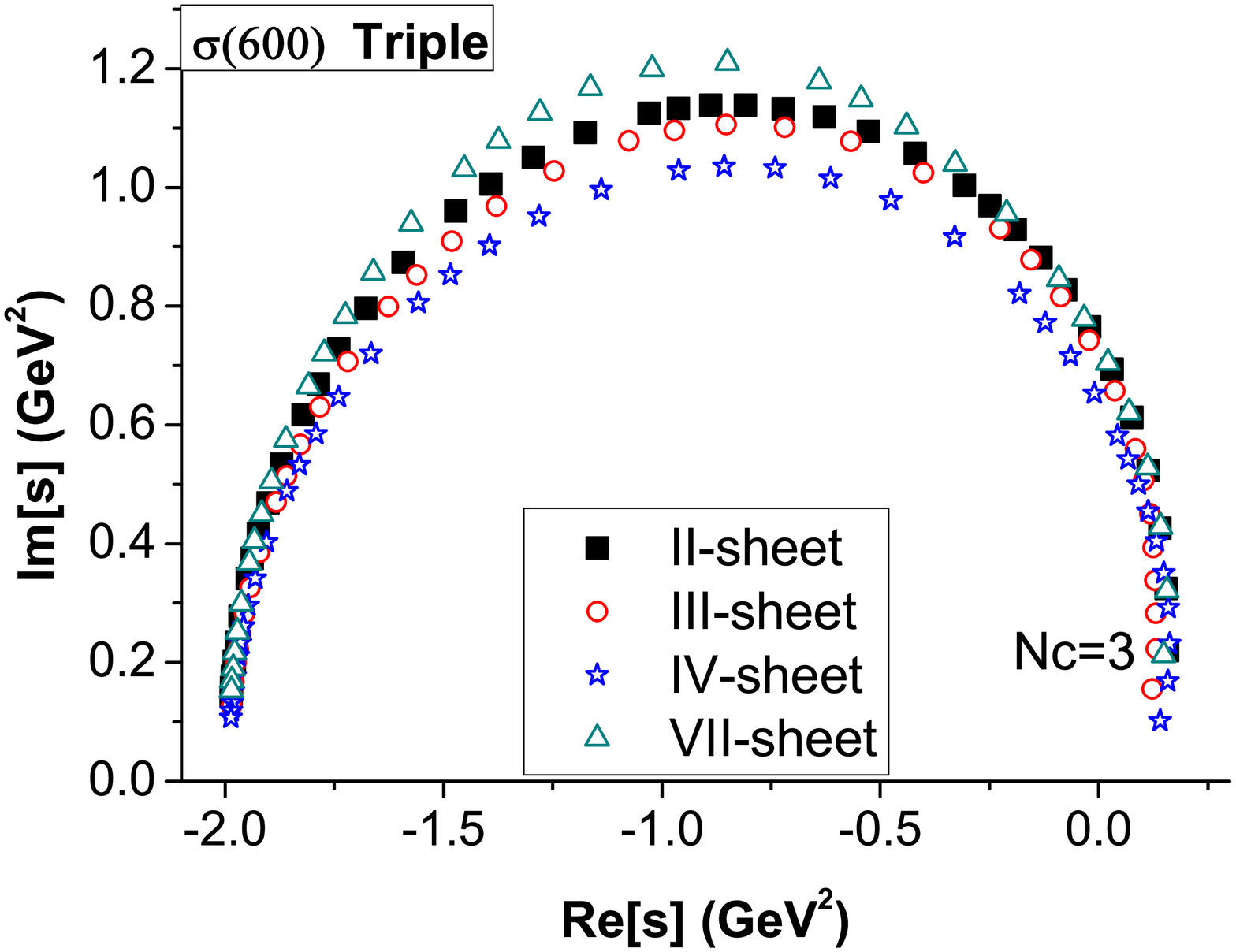}}
\end{minipage}
\caption{\label{Nc}$N_C$ trajectories.}
\end{figure}

With these LECs we get poles and their trajectories generated by
running  $N_c$. The $N_c$ trajectory of poles are plotted in
Fig.~\ref{Nc} c)--h).
The pole locations when $N_c=3$ are presented in
Table~\ref{pole}. For comparison, we also use these LECs  to plot
trajectories of $f_0(980)$, $\sigma$ in the old scheme of
Ref.~\cite{Dai11}. The plots are in Fig.~\ref{Nc}a) -- b).
The pole locations when $N_c$=3 are presented in
Table~\ref{pole}, too.
\begin{table}[h]
\begin{center}
 \begin{tabular}  {c c c c c}\hline
 Resonance       &        II          &        III        &  IV             &  VII          \\ \hline\hline
$2\times 2$      &  \                 &         \         &   \             &  -   \\
$f_0(980)$       &  $0.989-i0.005$    &         \         &   \             &  -   \\
$a_0(980)$       &        \           &   $0.707-i0.174$  &  $0.930-i0.442$ &  -   \\
$\sigma$         &  $0.441-i0.238$    &   $0.398+i0.130$  &   \             & -       \\
$\kappa$(700)    &  $0.746-i0.192$    &   $0.602+i0.263$  &   \             & -       \\
$K^{*}(892)$     &  $0.871-i0.020$    &   $0.903-i0.017$  &   \             & -         \\
$\rho(770)$      &  $0.760-i0.070$    &   $0.797-i0.058$  &   \             & -      \\
\hline
$3\times 3$      &  \                 &         \         &   \             &  \\
$f_0(980)$       &    $0.990-i0.004$  &   \               &  \              &  \\
$\sigma$         &  $0.462-i0.230$    &   $0.400-i0.194$  & $0.397-i0.128$  &   $0.453-i0.235$      \\
\hline
 \end{tabular}
 \caption{\label{pole}Resonance pole positions on $\sqrt{s}$ plane, in units of $\mathrm{GeV}$.}
 \end{center}
\end{table}
In the present case, the eight-sheets structure is rather
complicated for presenting results. Confined to simplified two
channel situation, it is found that the trajectories of $\rho$,
$K^*$, $a_0(980)$, $f_0(980)$ are qualitatively the same as our
previous work~\cite{Dai11}. It is worth pointing out that for
$\kappa$ and $\sigma$, a sheet III pole overlooked in
Ref.~\cite{Dai11} was found. We find that these two poles  satisfy
 the twin pole structure: poles in sheet II,III meet each other in the
real axis when $N_c=\infty$ (see Fig.~\ref{Nc}b) and e) ). On the
other side, for $f_0(980)$, even discussing in the triple channel
case, its trajectory is like what is found in Ref.~\cite{Dai11}: the
sheet II pole moves into upper half plane of sheet V from the lower
half of sheet II in the $\sqrt{s}$-plane, winding around the branch
point at the $K\bar{K}$ threshold (see Fig.~\ref{Nc}a), g)).
Eventually it will fall down to the positive real axis, between
$\pi\pi$ and $K\bar{K}$ threshold, with $no$ shadow poles
accompanied. The trajectory of $\sigma$ in the triple channel study
exhibits a clear quadruplet pole structure: there are four adjoint
poles in sheet II, III, IV, VII, and they will meet each other in
the negative real axis of the s complex plane when $N_c=\infty$ (see
Fig.~\ref{Nc}h) ).

\section{Conclusion}
In this paper, we have made a rather sophisticated  analysis on the
analyticity structure of Pad\'e amplitudes, with IJ=00 $\pi\pi-K\bar
K-\eta\eta$ channels included. Our`Breit-Wigner criteria'
previously proposed in Ref.~\cite{Dai11} is extended to
multi-channel case: A triple channel Breit-Wigner resonance should
appear as quadruplet poles and meet each other on the real axis when
$N_C=\infty$.
Our analysis confirm that
 $\rho$, $K^*$ are standard Breit-Wigner
particles due to their stable twin pole structure. For $a_0(980)$
one needs more precise data to get a better pole location when
$N_c=3$ and our analysis supports it to be a Breit-Wigner resonance.

For $f_0(980)$, we find that the qualitative picture and physical
conclusion one can draw from this analysis is very similar to our
previous results given in Ref.~\cite{Dai11}. It reveals that
$f_0(980)$ plays a very special role in the family of low lying
scalar mesons: it is most likely of $K\bar K$ nature. The meaning of
this paper is that we demonstrate such a distinguishing property of
$f_0(980)$ is stable against the number of thresholds it couples.

We also corrected the results on $\sigma$ and $\kappa$ made in
Ref.~\cite{Dai11}. Here we find that both $\sigma$ and $\kappa$
maintain a shadow pole structure, hence revealing their
Breit--Wigner origin, even though they behave very oddly in reality
and in the large $N_c$ world. Nevertheless the similarities between
two trajectories in sheet II and sheet III  even when $N_C$ is not
large suggests that they are dominantly single channel particles.

\section*{Acknowledgment}
This work is supported in part by  National Nature Science
Foundations of China under contract number 10925522  and
11021092.


\end{document}